\def\lb {\left[ }
\def\rb {\right] }
\def\lc {\left\{ }
\def\rc {\right\} }
\def\ra {\rangle }
\def\la {\langle }
\def\ni {\noindent}
\def\nn {\nonumber}

\def\rar {\rightarrow}
\def\lrar {\leftrightarrow}

\def\beq{\begin{equation}}
\def\eeq{\end{equation}}
\def\bea{\begin{eqnarray}}
\def\eea{\end{eqnarray}}

\def\cL {{\cal{L}}}

\def\cO{{\cal{O}}}

\def\d{\delta}
\def\D {\Delta}
\def\e{\epsilon}

\def\m{\mu}

\def\O {\Omega}
\def\p {\pi}
\def\r{\rho}
\def\s{\sigma}

\def\ub {\bar u}

\def\sm {\!-\!}
\def\st {\!\times \!}
\def\cd {\!\cdot\!}
\def\sa {\!\rar\!}

\def\br {\mbox{\boldmath $r$}}

\def\bsig {\mbox{\boldmath $\sigma$}}
\def\btau {\mbox{\boldmath $\tau$}}

\def\bnb {\mbox{\boldmath $\nabla$}}

\documentclass{ws-mpla}

\begin{document}

\markboth{M. R. Robilotta}
{Nuclear Interactions: the Chiral Picture}

\catchline{}{}{}{}{}

\title{NUCLEAR INTERACTIONS: THE CHIRAL PICTURE}

\author{M.R. Robilotta}

\address{Instituto de F\'{i}sica, Universidade de S\~ao Paulo,
S\~ao Paulo, SP, Brazil}

\maketitle

\pub{Received (Day Month Year)}{Revised (Day Month Year)}

\begin{abstract}
Chiral expansions of the two-pion exchange components of both two- and 
three-nucleon forces are reviewed and a discussion is made of the 
predicted pattern of hierarchies.
The strength of the scalar-isoscalar central potential is found to be
too large and to defy expectations from the symmetry.
The causes of this effect can be understood by studying the
nucleon scalar form factor.

\end{abstract}

\ccode{PACS Nos.: 13.75.Gx, 21.30.Fe}

\section{CHIRAL SYMMETRY}
\label{s1}

\begin{figure}[h]
\begin{center}
\includegraphics[width=0.7\columnwidth,angle=0]{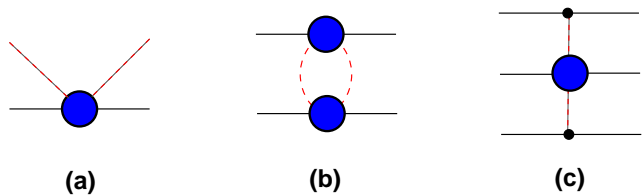}
\vspace{-3mm}
\caption{Free $\p N$ amplitude (a)
and two-pion exchange two-body (b) and three-body (c) potentials.} 
\label{F1}
\end{center}
\end{figure}

The outer components of nuclear forces are dominated by pion-exchanges
and involve just a few basic subamplitudes, describing pion interactions
with either nucleons or other pions.
The simplest process $N \sa \p N$, corresponding to the emission or 
absorption of a single pion by a nucleon, is rather well understood and 
gives rise to the one-pion exchange  $N\!N$ potential ($OPEP$).
The scattering reaction $\p N \sa \p N$ comes next and determines both 
the very important two-pion exchange term in the $N\!N$ force 
and the leading three-body interaction, as shown in Fig.\ref{F1}.

\vspace{-4mm}

\vspace{-5mm}

The theoretical understanding of the $\p N$ amplitude proved
to be very challenging and a suitable description was only produced 
by means of chiral symmetry.
This framework provides a  natural explanation for the observed 
smallness of $\p N$ scattering lengths and plays a fundamental
role in Nuclear Physics.
Nowadays, the use of chiral symmetry in low-energy pion-interactions
is justified by $QCD$.

The small masses of the quarks $u$ and $d$, treated as perturbations 
in a chiral symmetric lagrangian, give rise to a well defined chiral 
perturbation theory (ChPT).
Hadronic amplitudes are then expanded in terms of a typical 
scale $q$, set by either pion four-momenta or nucleon three-momenta, 
such that $q\ll 1$ GeV.
This procedure is rigorous and many results have the status of {\em theorems}.
In general, these theorems are written as power series 
in the scale $q$ and involve both {\em leading order terms} and 
{\em chiral corrections}.
The former can usually be derived from tree diagrams, whereas the
latter require the inclusion of pion loops and are the main object 
of ChPT. 
At each order, predictions for a given process must be unique and the 
inclusion of corrections cannot change already existing leading terms.

The relationship between chiral expansions of the $\p N$ amplitude and of 
two-pion exchange $(TPE)$ nuclear forces is discussed in the sequence.
For the $\p N$ amplitude, tree diagrams yield $\cO(q, q^2)$ terms 
and corrections up to $\cO(q^4)$ have already been evaluated, by means of both 
covariant\cite{BL}(CF) and heavy baryon\cite{FM}(HBF) formalisms. 
In the case of the $N\!N$ potential, the leading term is $\cO(q^0)$ and
given by the $OPEP$.
The tree-level $\p N$ amplitude yields $TPE$ contributions at $\cO(q^2,q^3)$
and corrections at $\cO(q^4)$ are available, based on both 
HBF\cite{HB} and CF\cite{HR,HRR}.
Tree-level $\p N$ results also determine the leading $\cO(q^3)$ three-body
force and partial corrections at $\cO(q^4)$ begin to be derived 
\cite{IR07,E3NP}.
As this discussion suggests, $\cO(q^4)$ corrections to both
two- and three-nucleon forces require just the $\cO(q^3)$
$\p N$ amplitude.

\begin{figure}[h]
\begin{center}
\includegraphics[width=0.8\columnwidth,angle=0]{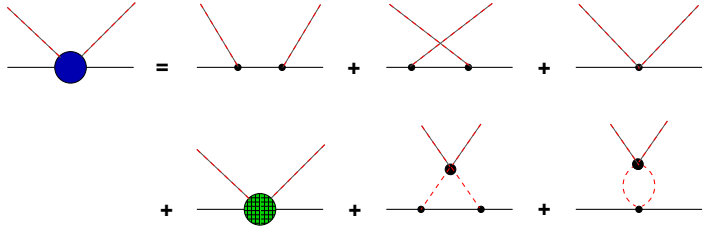}
\vspace{-4mm}
\caption{Representation of the $\p N$ amplitude at $\cO(q^3)$.} 
\label{F2}
\end{center}
\end{figure}

The full empirical content of the $\p N$ amplitude cannot be predicted
by chiral symmetry alone. 
Experimental information at low energies is usually encoded into the 
subthreshold coefficients introduced by H\"ohler and collaborators\cite{H83}
which can, if needed, be translated into the low-energy contants (LECs)
of chiral lagrangians. 
Therefore, in order to construct a $\cO(q^3)$ $\p N$ amplitude,
one uses chiral symmetry supplemented by 
subthreshold information, as indicated in Fig. \ref{F2}.
The first two diagrams correspond to the nucleon pole, 
whereas the other ones represent a smooth background.
The third graph reproduces the Weinberg-Tomozawa contact interaction,
the fourth one summarizes LEC contributions and the last two describe 
medium range pion-cloud effects.

\vspace{-5mm}

\section{TWO-BODY POTENTIAL}
\label{s2}

\begin{figure}[h]
\begin{center}
\includegraphics[width=0.7\columnwidth,angle=0]{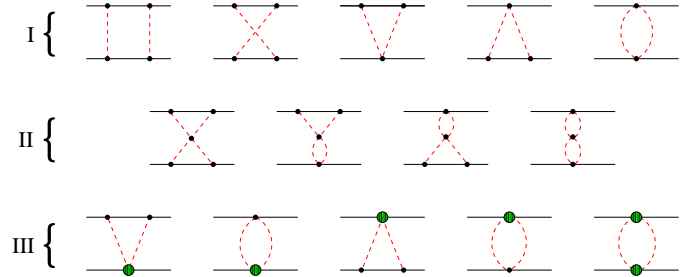}
\caption{Dynamical structure of the two-pion exchange potential.} 
\label{F3}
\end{center}      
\end{figure}

With the purpose of discussing the problem of predicted 
$\times$ observed chiral hierarchies, in this section we review 
briefly results obtained by our goup\cite{HR,HRR} for the 
$TPE$-$N\!N$ potential at $\cO(q^4)$.
This component is determined by the three families of diagrams
shown in Fig. \ref{F3}. 
Family $I$ begins at $\cO(q^2)$ and implements the minimal realization of 
chiral symmetry\cite{RR94},
whereas family $I\!I$ depends on $\p\p$ correlations and is $\cO(q^4)$.
They involve only the constants $g_A$ and $ f_\pi$
and all dependence on the LECs is concentrated in family $I\!I\!I$,
which begins at $\cO(q^3)$.

\vspace{-2mm}

\vspace{-5mm}

As far as chiral orders of magnitude are concerned, on finds that 
the various components of the force begin as follows\cite{HRR}:
$\cO(q^2)\rar V_{SS}^+,  V_T^+, V_C^-$ and 
$\cO(q^3)\rar V_C^+, V_{LS}^+, V_{LS}^-, V_{SS}^-, V_T^-$,
where the superscripts $(+)$ and $(-)$ refer to terms proportional 
to either the identity or $\btau^{(1)}\cd \btau^{(2)}$ in isospin space.
An interesting feature of these results is that the role played by 
family $I\!I$ is completely irrelevant.
On the other hand, family $I$ dominates almost completely the components
$V_{LS}^+$, $V_T^+$, $V_{SS}^+$ and $V_C^-$, 
whereas family $I\!I\!I$ does the same for $V_C^+$, $V_T^-$and $V_{SS}^-$.

\vspace{-4mm}
\begin{figure}[h]
\begin{center}
\includegraphics[width=0.49\columnwidth,angle=0]{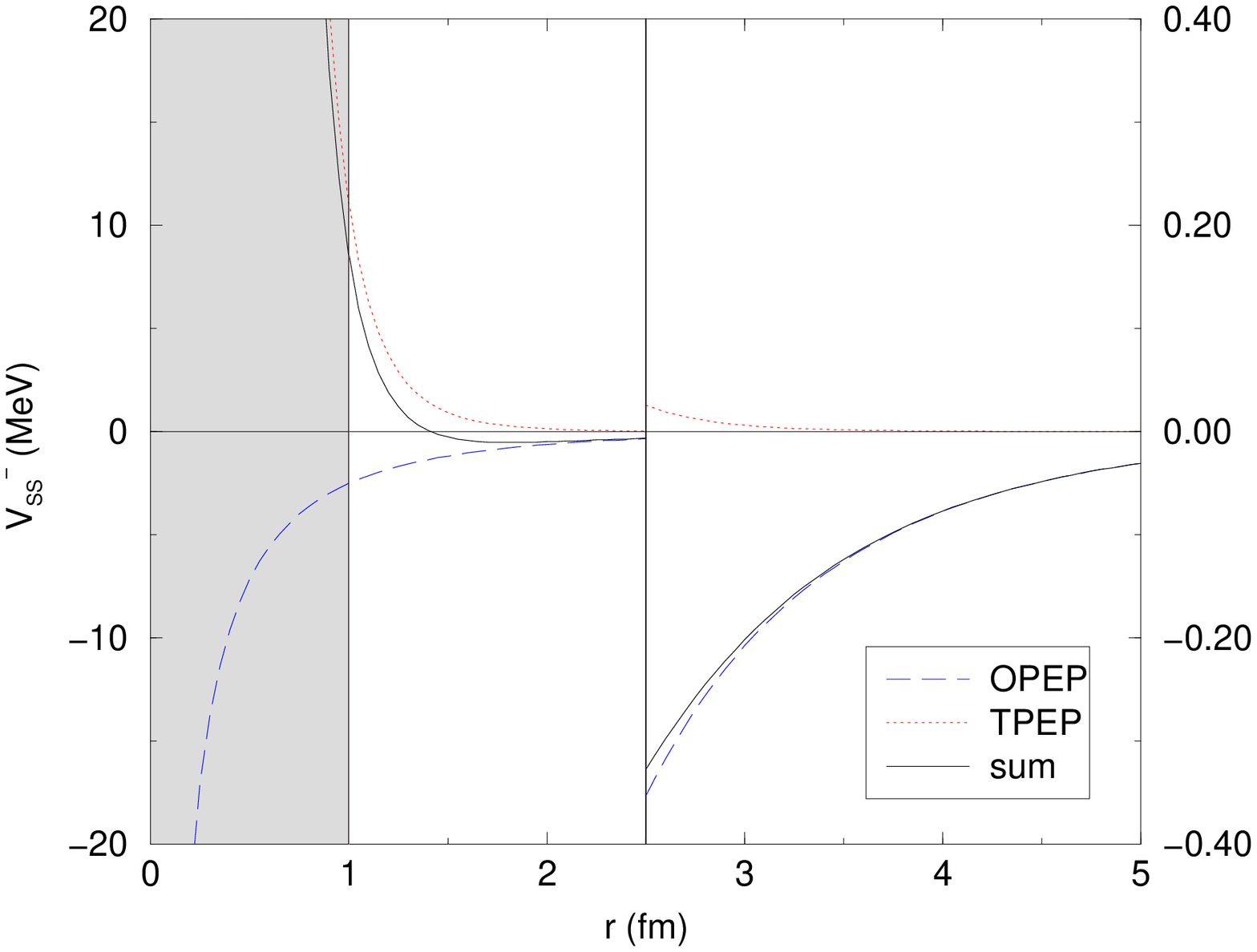}
\includegraphics[width=0.49\columnwidth,angle=0]{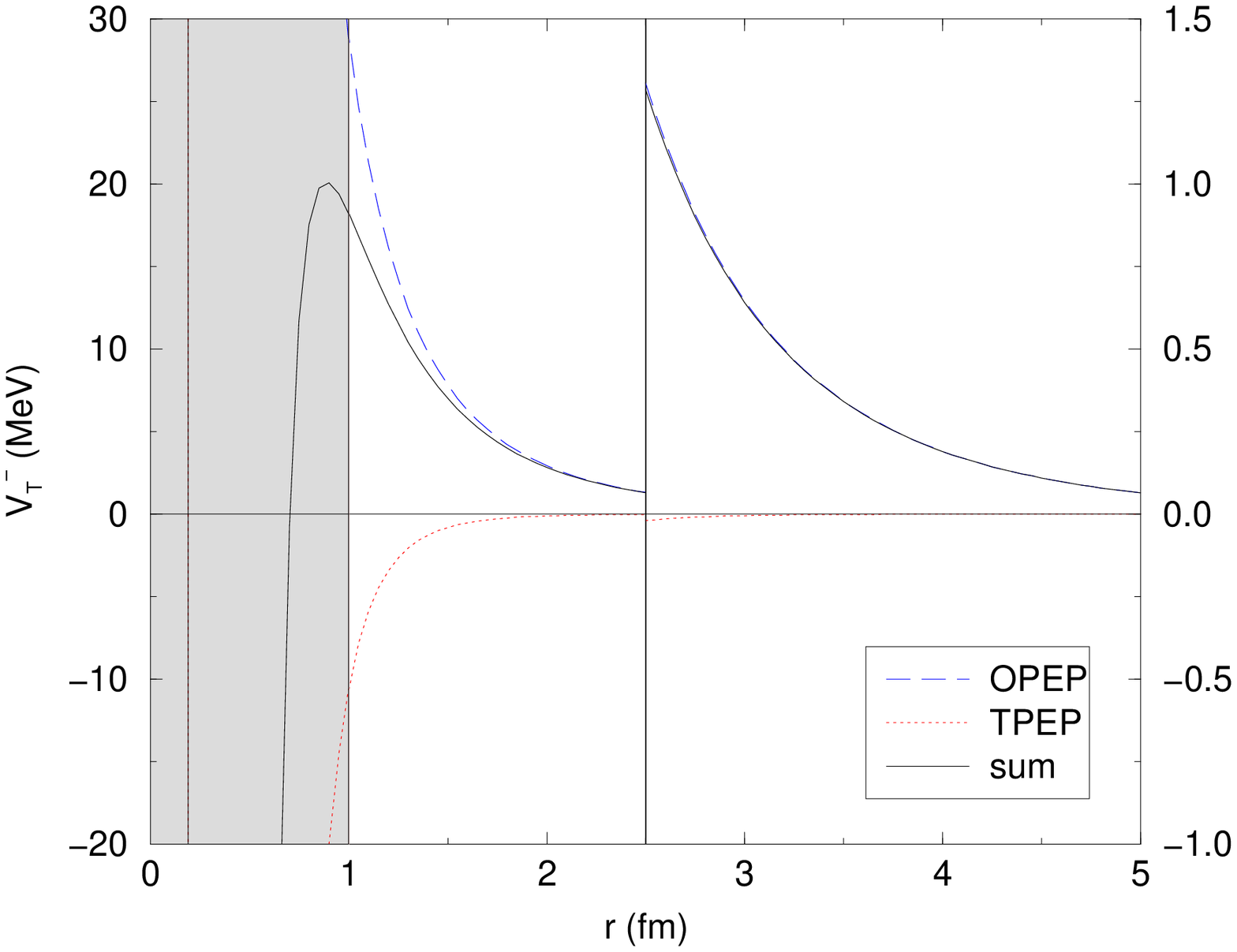}
\vspace{-3mm}
\caption{$OPEP$ and $TPEP$ contributions to 
spin-spin (left) and tensor (right) NUisovector components.} 
\label{F4}
\end{center}      
\end{figure}

\vspace{-5mm}

The relationship between the $OPEP[=\cO(q^0)]$ and $TPEP[=\cO(q^3)]$ 
contributions to the $V_{SS}^-$ and $V_T^-$ profile functions is 
shown in Fig. \ref{F4}, where it is possible to see that the chiral
hierarchy is respected.

\begin{figure}[h]
\begin{center}
\hspace*{-4mm}
\includegraphics[width=0.50\columnwidth,angle=0]{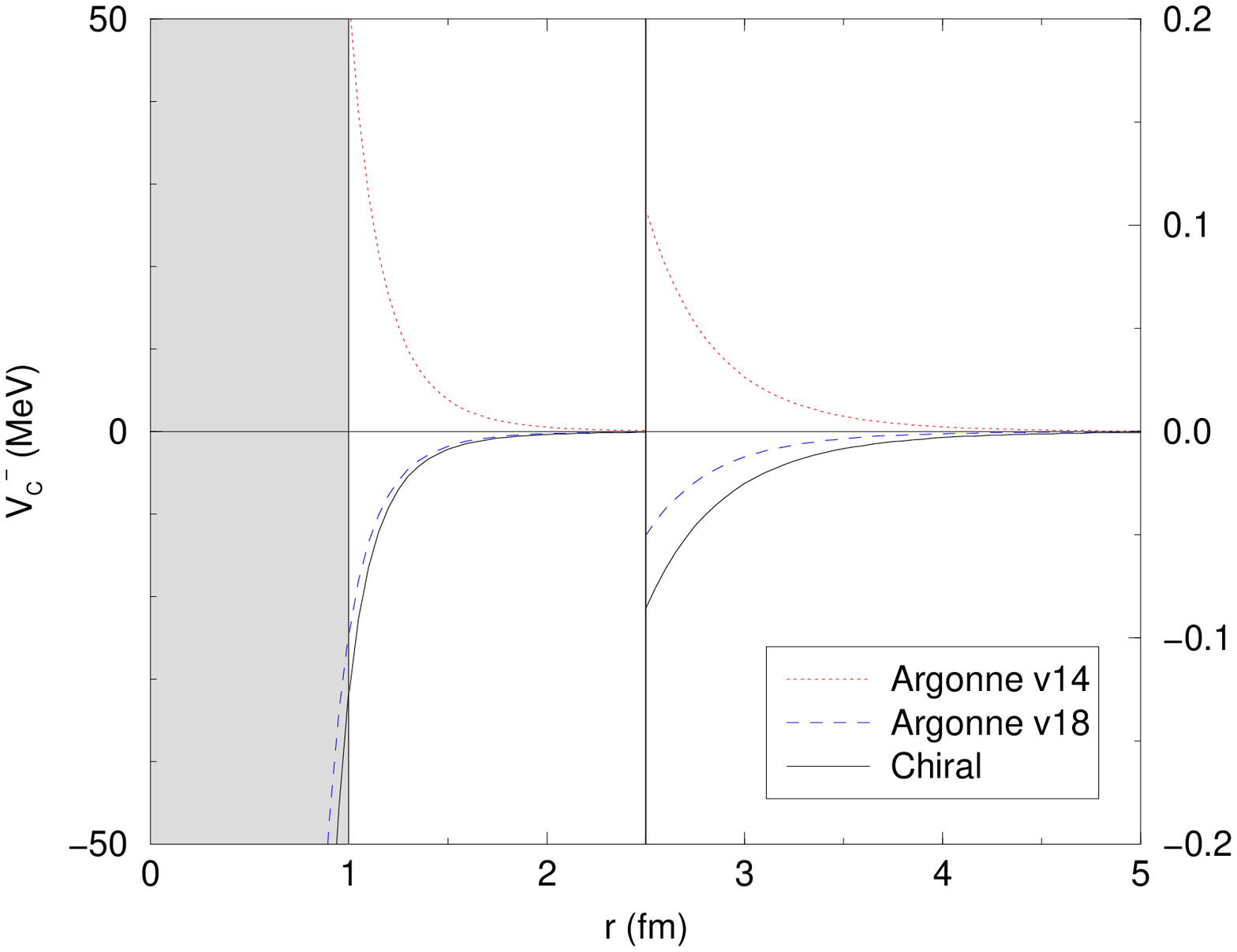}
\includegraphics[width=0.50\columnwidth,angle=0]{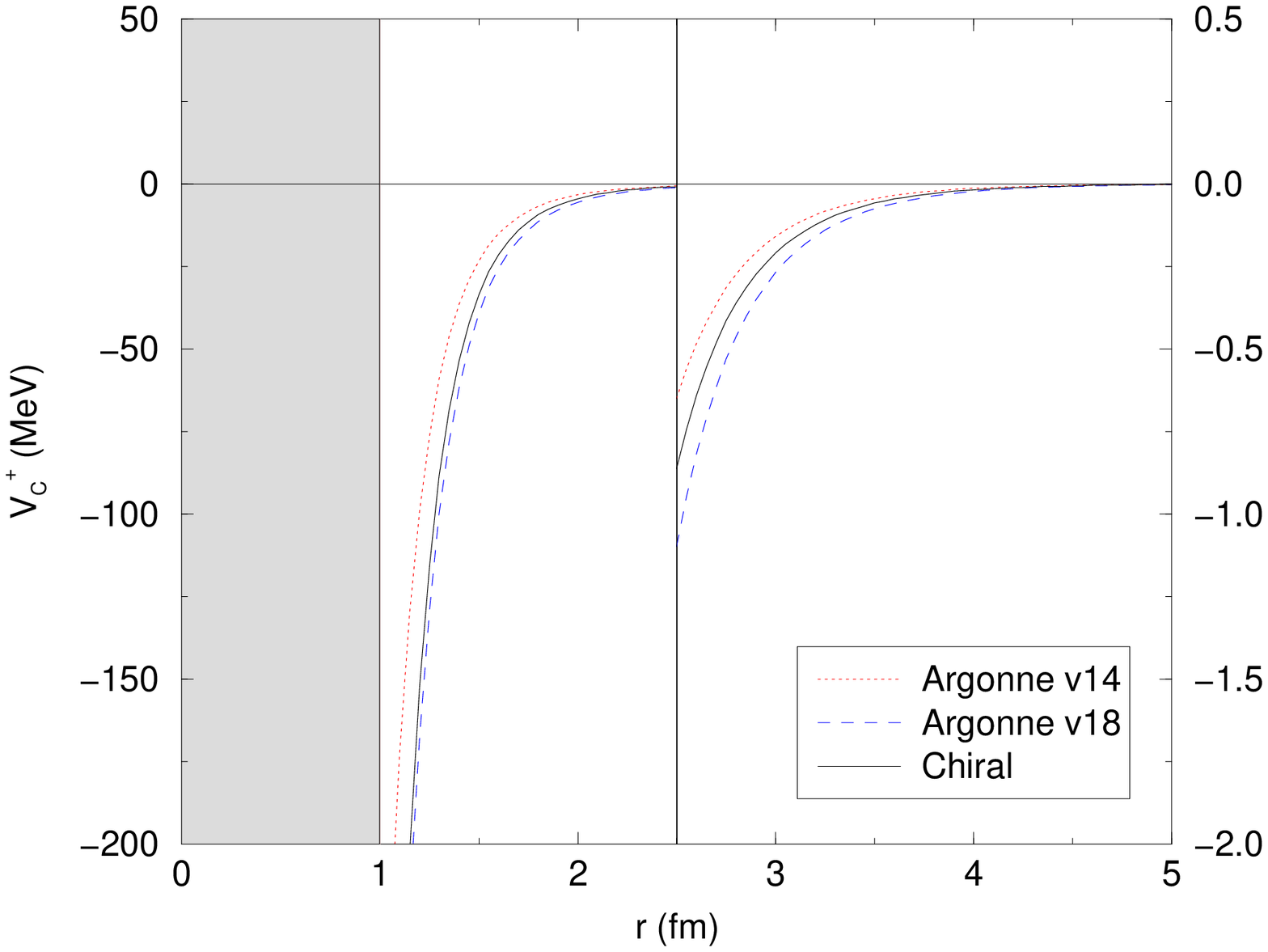}
\vspace{-2mm}
\caption{Isospin odd (left) and even (right) central components of the 
two-pion exchange potential.} 
\label{F5}
\end{center}      
\end{figure}

In Fig. \ref{F5}, the two central components $V_C^-[=\cO(q^2)]$ and 
$V_C^+[=\cO(q^3)]$ are displayed side by side and two features 
are to be noted.
The first one concerns the favorable comparison with the empirical
Argonne\cite{Arg} potentials in both cases.
The second one is that $|V_C^+| \sim 10 \, |V_C^-|$ in  regions of physical
interest, defying strongly the predicted chiral hierarchy.
This problem will be further discussed in the sequence.

\vspace{-3mm}

\vspace{-4mm}

Violations of the chiral hierarchy are also present in the 
{\em drift potential}\cite{Rdrift}, which corresponds to kinematical
corrections due to the fact that the two-body center of mass is allowed to 
drift inside a larger system. 
In terms of Jacobi coordinates, it is represented by the operator
\bea
V(r)^\pm  =  \left. V(r)^\pm \rb_{cm} + V_D^\pm \, \O_{D}
\;\;\;\;\;\;\;\; \lrar \;\;\;\;\;\;\;\;
\O_D = \frac{1}{4\sqrt{3}}\, (\bsig^{(1)}\sm \bsig^{(2)}) \cd \br \st \;,
(-i \bnb^{^{\!\!\!\!\!\!\!\!^\leftrightarrow}}_\r)\;.
\nn
\eea

\begin{figure}[h]
\begin{center}
\includegraphics[width=0.7\columnwidth,angle=0]{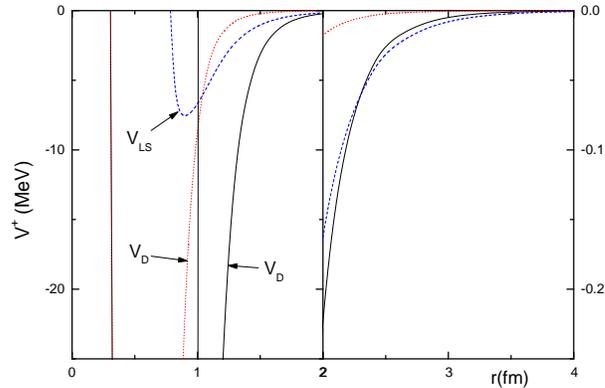}
\vspace{-35mm}
\caption{Isospin even drift (full and dotted lines)
and spin-orbit (dashed line) potentials.} 
\label{F6}
\end{center}      
\end{figure}

The profile function $V_D^+$ together with $V_{LS}^+$,  are 
displayed in Fig. \ref{F6}.
Drift corrections begin at $\cO(q^4)$ and, in principle, should be smaller 
than the spin-orbit terms, which begin at $\cO(q^3)$.
However, in this channel, the hierarchy is again not respected.

\vspace{20mm}

\section{THREE-BODY POTENTIAL}
\label{s3}

The leading term in the three-nucleon potential, known as $TPE$-$3NP$,
has long range and corresponds to the process shown in fig.1c,
in which a pion is emitted by one of the nucleons, scattered by a second one, 
and absorbed by the last nucleon.
Is this case, the intermediate $\p N$ amplitude, which is $\cO(q)$ for 
free pions, becomes $\cO(q^2)$ and the three-body force begins at 
$\cO(q^3)$.
The first modern version of this component of the force was produced by 
Fujita and Miyazawa\cite{F-M}, its chiral structure has bee much debated
since the seventies\cite{3NP} and, nowadays, a sort o consensus has been 
reached about its form\cite{Rtokyo}. 
The leading $TPE$-$3NP$ has a generic structure given by
\bea
&& V_L(123) = -\,\frac{\m}{(4\p)^2}  
\lc \d_{ab} \lb a \, \m - b \, \m^3 \, \bnb_{12} \cd \bnb_{23} \rb 
+ d\, \m^3 \; i\,\e_{bac} \tau_c^{(2)}\;
i \, \bsig^{(2)} \cd \bnb_{12} \times \bnb_{23}\rc
\nn\\
&& \;\;\;\;\; \times 
\lb (g_A \, \m/2 \,f_\p)\;\tau_a^{(1)} \;\bsig^{(1)} \cd \bnb_{12} \rb\;
\lb (g_A \, \m/2 \,f_\p)\;\tau_b^{(3)} \;\bsig^{(3)} \cd \bnb_{23} \rb\;
Y(x_{12}) \; Y(x_{23}) \;,
\nn
\eea

\ni
where $\m$ is the pion mass and $a$, $b$ and $d$ are strength parameters,
determined by either LECs or subhtreshold coefficients. 

The evaluation of $\cO(q^4)$ corrections requires the inclusion of single
loop effects and is associated with a large number of 
diagrams, which are being calculated by Epelbaum and 
collaborators\cite{E3NP}.
In order to produce a feeling for the structure of these corrections,
we discuss a particular set of processes belonging to the 
$TPE$-$3NP$ class, considered recently\cite{IR07}.
Full results involve expressions which are too long and cumbersome
to be displayed here.
However, their main qualitative features can be summarized in the
structure  
$V(123)=V_L(123)+ [V_{\d L}(123) + \d V(123)]$, 
where $V_L$ is the leading term shown above and the factors within 
square brackets are ChPT corrections.
The function $V_{\d L}$ can be obtained directly from $V_L$, by replacing 
$(a,b,c) \rightarrow (\d a,\d b, \d c)$, where 
the $\d s$ indicate changes smaller than $10\%$.
This part of the ChPT correction corresponds just to shifts in the 
parameters of the leading component.
The term $\d V(123)$, on the other hand, represents effects associated 
with new mathematical functions involving both non-local operators 
and complicated propagators containg loop integrals,
in place of the Yukawa functions.
The strengths  of these new functions are determined by a new set of 
parameters $e_i$, which are also typically about $10\%$ of the 
leading ones.

In summary, ChPT gives rise both to small changes in already existing 
coefficients and to the appearance of many new mathematical structures.
The latter are the most interesting ones, since they may be instrumental
in explaining effects such as the $A_y$ puzzle.

\section{THE CHIRAL PICTURE}
\label{s4}

Chiral symmetry has already been applied to about 20 components 
of nuclear forces, allowing a comprehensive picture to be assessed.
According to ChPT, the various effects begin to appear at different
orders and the predicted hierarchy is displayed in the table below.

\begin{table}[h]
\begin{center}
\begin{tabular} {|c|ccc|}
\hline
beginning	  	& TWO-BODY & TWO-BODY	& THREE-BODY 	   \\ 
				& $OPEP$   & $TPEP$		& $TPEP$		   \\ \hline 
$\cO(q^0)$		& $V_T^-, V_{SS}^-$	&&    	  	   \\[1mm] \hline
$\cO(q^2)$		& $V_D^-$ & $V_C^-; V_T^+, V_{SS}^+$ &	   \\[1mm]\hline
$\cO(q^3)$		&& $V_{LS}^-, V_T^-, V_{SS}^-; V_C^+, V_{LS}^+$ 
& $d; a, b$ 	\\[1mm] \hline
$\cO(q^4)$		&& $V_D^-; V_Q^+, V_D^+$ & $ e_i $ \\[1mm] \hline
\end{tabular}
\end{center}
\end{table}

In Ref.\refcite{HRR}, the relative importance of $O(q^2)$, $O(q^3)$ 
and $O(q^4)$ terms in each component of the $TPEP$-$N\!NP$ has been studied.
In general, convergence at distances of physical interest is 
satisfactory, except for $V_C^+$, where the ratio between 
$\cO(q^4)$ and $\cO(q^3)$ contributions 
is larger than $0.5$ for distances smaller than $2.5$ fm.

As far as the relative sizes of the various dynamical effects are
concerned, one finds strong violations of the predicted hierarchy
when one compares $V_C^+$ with $V_C^-$ and $V_D^+$ with $V_{LS}^+$,
as discussed above. 
It is interesting to note that, in both cases,
the unexpected enhancements occur in the isoscalar sector. 
The numerical explanation for this behavior is that some of the LECs 
used in the calculation are large and
generated dynamically  by delta intermediate states.
However, it is also possible that perturbation theory may not apply 
to isoscalar interactions at intermediate distances.
This aspect of the problem is explored in the next section.

\section{SCALAR FORM FACTOR}
\label{s5}

The structure of $V_C^+$ was  scrutinized in Ref.\refcite{HRR}
and  found to be heavily dominated by a term of the form
\bea
V_C^+(r) \sim -\, (4/f_\p^2)\;\lb (c_3 - 2c_1) - c_3 \; \bnb^2/2 \rb \;
\tilde{\s}_{N_N}(r)\;,
\nn
\eea
where the $c_i$ are LECs and $\tilde{\s}_{N_N}$ is the leading contribution 
from the pion cloud to the nucleon scalar form factor.
This close relationship between $\tilde{\s}_{N_N}$ and $V_C^+$ indicates 
that the study of the former can shed light into the properties of the 
latter.

The nucleon scalar form factor is defined as 
\bea
\la N(p') | \sm \cL_{sb}\, | N(p) \ra = \s_N(t) \; \ub(p')\; u(p) \;,
\nn
\eea

\ni
where $\cL_{sb}$ is the symmetry breaking lagrangian. 
It has already been expanded\cite{BL} up to $\cO(q^4)$ and receives its leading 
$\cO(q^2)$ contribution from a tree diagram associated with the LEC $c_1$.
Corrections at $\cO(q^3)$ and $\cO(q^4)$ are produced by two triangle diagrams, 
involving nucleon and delta intermediate states.
In configuration space\cite{sigma}, the scalar form factor is denoted
by $\tilde{\s}$ and one writes 
\bea
\tilde{\s}_N(\br) = - 4\, c_1\, \m^2\, \delta^3(\br) + 
\tilde{\s}_{N_N}(r) + \tilde{\s}_{N_\D}(r) \;,
\nn
\eea

\ni
where $\tilde{\s}_{N_N}$ and $\tilde{\s}_{N_\D}$ are the finite-range 
triangle contributions. 

\begin{figure}[t]
\begin{center}
\hspace*{-4mm}
\includegraphics[width=0.35\columnwidth,angle=-90]{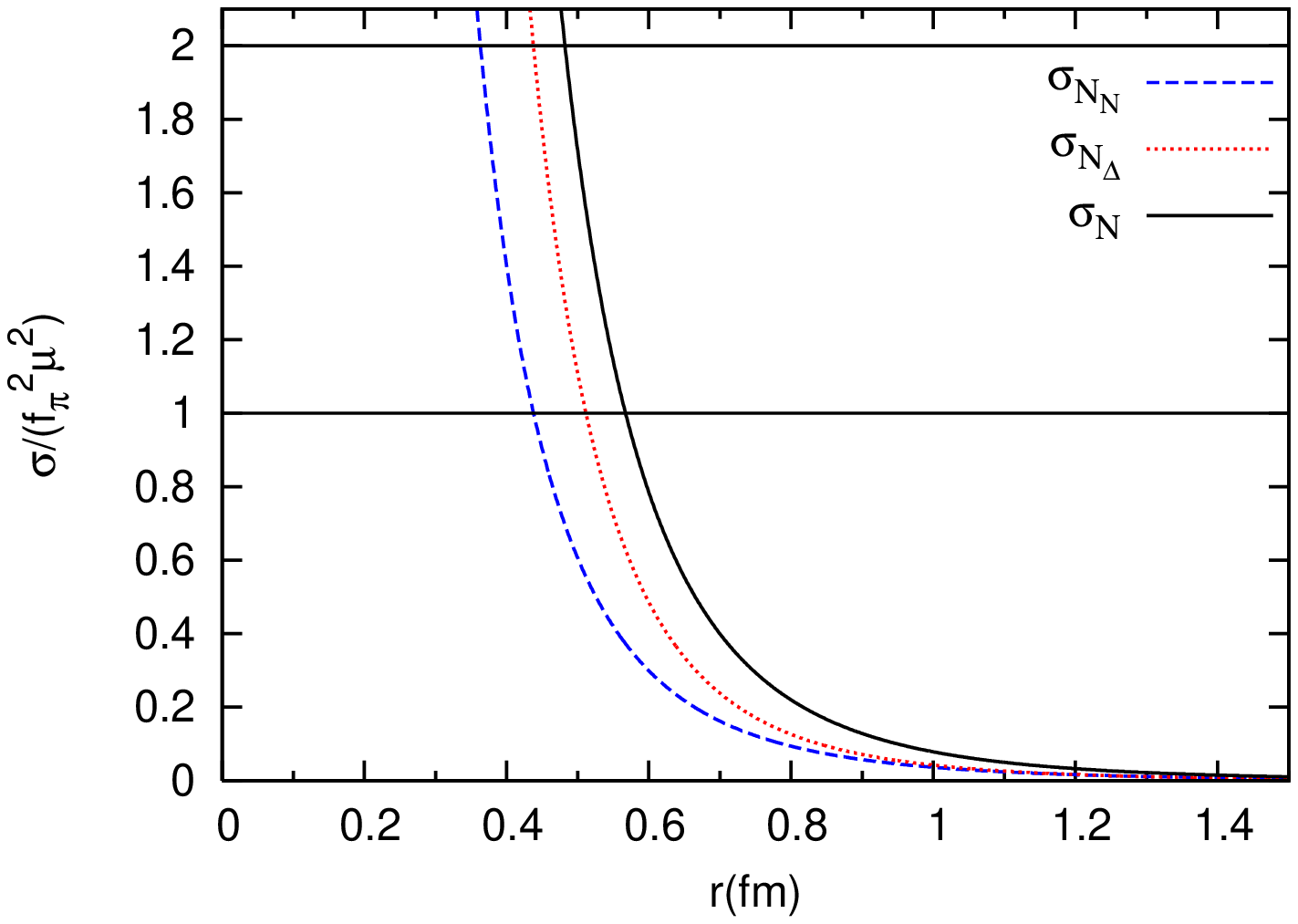}
\includegraphics[width=0.35\columnwidth,angle=-90]{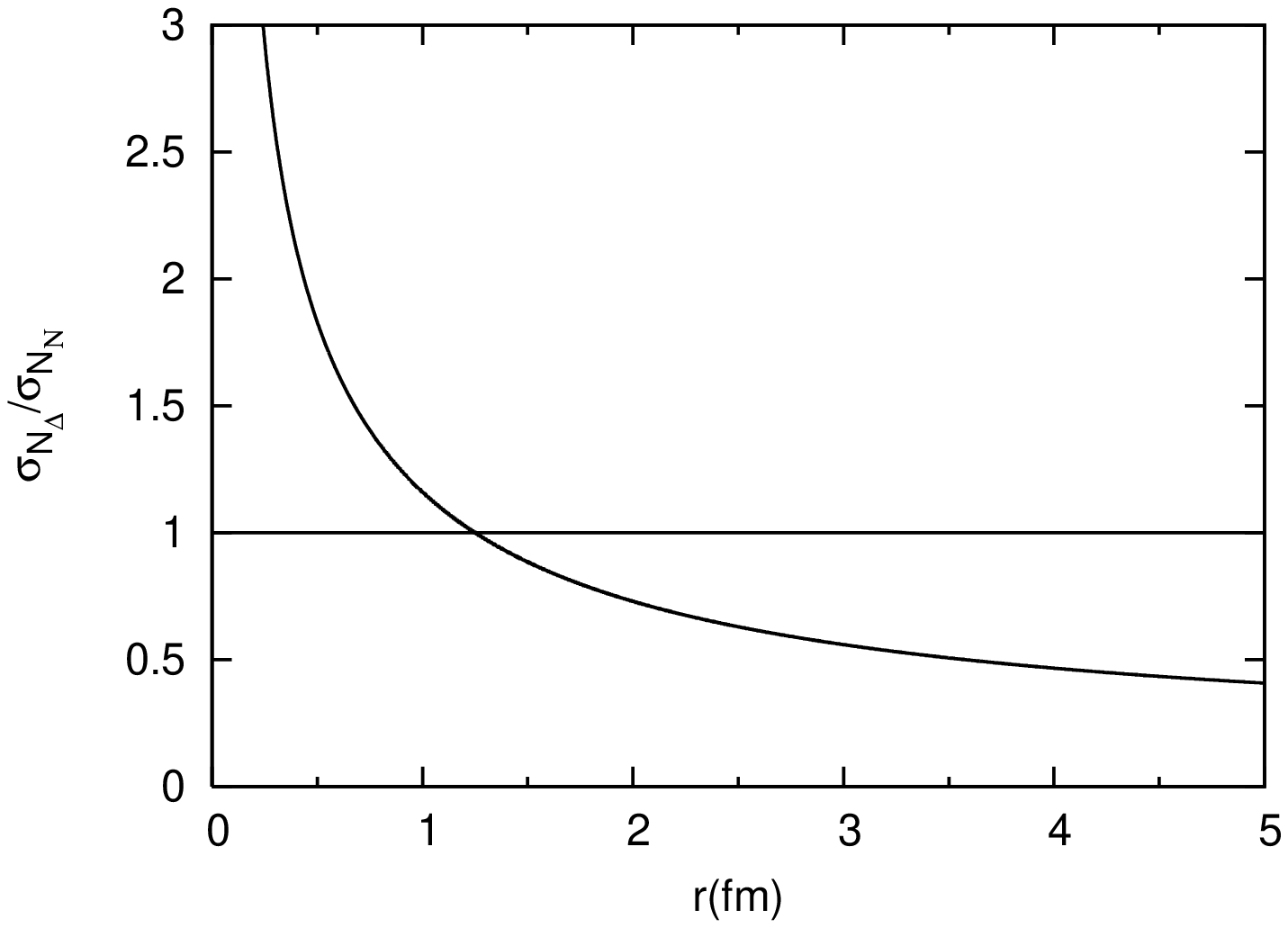}
\caption{Ratios $\tilde{\s}_N(r)/(\m^2 f_\p^2)=(1-\cos\theta)$ (left)
and $\tilde{\s}_{N_\D}(r)/\tilde{\s}_{N_N}(r)$ (right) as functions of 
the distance $r$.}
\label{F7}
\vspace{-3mm}
\end{center}      
\end{figure}

The symmetry breaking lagrangian can be expressed in terms of the chiral angle 
$\theta$ as $\cL_{sb} = f_\p^2 \, \m^2 \,(\cos\theta-1)$.
The ratio $\tilde{\s}_N(r)/(\m^2 f_\p^2)=(1-\cos\theta)$ describes the 
density of the $q\bar{q}$ condensate around the nucleon and is 
displayed in Fig.~\ref{F7}.
One notes that it vanishes at large distances and increases monotonically 
as one approaches the center.
This means that the function $\tilde{\s}_N(r)$ becomes meaningless beyond 
a critical radius $R$, corresponding to $\theta = \p/2$,
since the physical interpretation of the quark condensate 
requires the condition $q\bar{q}>0$.
In Ref. \refcite{sigma}, the condensate was assumed to no longer exist in 
the region $r<R$ and the $\p N$ sigma-term was evaluated using the expression
\bea
\s_N = \frac{4}{3} \p R^3 \;f_\p^2 \m^2
+ 4\p \int_{R}^\infty dr\;r^2\;\tilde{\s}_N(\br)\;.
\nn
\eea

This procedure yields 43 MeV$<\s_N<\;$49 MeV, depending on the value 
adopted for the $\p N\Delta$ coupling constant, in agreement  
with the empirical value 45$\pm 8$ MeV.
This picture of the nucleon scalar form factor is sound 
and can be used to gain insight about $V_C^+$.

Inspecting Fig. \ref{F7} (right), one learns that the hierarchy predicted 
by ChPT is subverted for distances smaller than 1.5 fm, since the 
$\cO(q^4)$ delta becomes more important than the $\cO(q^3)$ nucleon.
On the other hand, the good prediction obtained for the nucleon $\s$-term 
(and also for the $\D$ $\s$-term\cite{sigma}) 
indicates that the functions $\tilde{\s}_{N_N}(r)$ and $\tilde{\s}_{N_\D}(r)$
can be trusted up to the critical radius $R\sim$~0.6~fm.
Just outside this radius, the chiral angle is close to $\p/2$,
indicating that the pion cloud is non-perturbative in that region.
This picture is supported by Fig. \ref{F5} (right) since, 
at least up to 1 fm, the prediction for $V_C^+$ agrees well with 
the Argonne phenomenological potentials.
This leads to our main conclusion, namely that the range of validity of 
calculations based on nucleon and delta intermediate states is wider that 
that predicted by ChPT.

\section*{Acknowledgments}

It was a great pleasure participating in the Chiral 2007 meeting
and I would like to thank the organizers for the very nice conference,
for the warm and friendly hospitality, and for supporting my 
stay in Osaka.


\end{document}